\documentclass[longauth]{aa} 
\usepackage{graphicx}
\usepackage{lscape}
\begin{document}
   \title{Upper Limits from HESS AGN Observations in 2005$-$2007}


\author{F. Aharonian\inst{1,13}
 \and A.G.~Akhperjanian \inst{2}
 \and U.~Barres de Almeida \inst{8} \thanks{supported by CAPES Foundation, Ministry of Education of Brazil}
 \and A.R.~Bazer-Bachi \inst{3}
 \and B.~Behera \inst{14}
 \and M.~Beilicke \inst{4}
 \and W.~Benbow \inst{1}
 \and K.~Bernl\"ohr \inst{1,5}
 \and C.~Boisson \inst{6}
 \and O.~Bolz \inst{1}
% ...
}

\offprints{Wystan.Benbow@mpi-hd.mpg.de}

\institute{
Max-Planck-Institut f\"ur Kernphysik, P.O. Box 103980, D 69029
Heidelberg, Germany
\and
 Yerevan Physics Institute, 2 Alikhanian Brothers St., 375036 Yerevan,
Armenia
\and
Centre d'Etude Spatiale des Rayonnements, CNRS/UPS, 9 av. du Colonel Roche, BP
4346, F-31029 Toulouse Cedex 4, France
\and
Universit\"at Hamburg, Institut f\"ur Experimentalphysik, Luruper Chaussee
149, D 22761 Hamburg, Germany
\and
Institut f\"ur Physik, Humboldt-Universit\"at zu Berlin, Newtonstr. 15,
D 12489 Berlin, Germany
\and
LUTH, Observatoire de Paris, CNRS, Universit\'e Paris Diderot, 5 Place Jules Janssen, 92190 Meudon, 
France
\and
DAPNIA/DSM/CEA, CE Saclay, F-91191
Gif-sur-Yvette, Cedex, France
\and
University of Durham, Department of Physics, South Road, Durham DH1 3LE,
U.K.
\and
Unit for Space Physics, North-West University, Potchefstroom 2520,
    South Africa
\and
Laboratoire Leprince-Ringuet, Ecole Polytechnique, CNRS/IN2P3,
 F-91128 Palaiseau, France
\and 
Laboratoire d'Annecy-le-Vieux de Physique des Particules, CNRS/IN2P3,
9 Chemin de Bellevue - BP 110 F-74941 Annecy-le-Vieux Cedex, France
\and
Astroparticule et Cosmologie (APC), CNRS, Universite Paris 7 Denis Diderot,
10, rue Alice Domon et Leonie Duquet, F-75205 Paris Cedex 13, France
\thanks{UMR 7164 (CNRS, Universit\'e Paris VII, CEA, Observatoire de Paris)}
\and
Dublin Institute for Advanced Studies, 5 Merrion Square, Dublin 2,
Ireland
\and
Landessternwarte, Universit\"at Heidelberg, K\"onigstuhl, D 69117 Heidelberg, Germany
% ...
}

   \date{Received 4 September 2007 / Accepted 6 November 2007}

 
  \abstract
   {}
   {Very high energy (VHE; E$>$100 GeV) $\gamma$-ray studies 
were performed for 18 active galactic nuclei (AGN) from
a variety of AGN classes.}
   {VHE observations of a sample of 14 AGN, considered
candidate VHE emitters, were made with the 
High Energy Stereoscopic System (HESS)
between January 2005 and July 2007. Large-zenith-angle
observations of three northern
AGN (Mkn\,421, Mkn\,501, 1ES\,1218+304),
known to emit VHE $\gamma$-rays, were also 
performed in order to sample their spectral
energy distributions (SEDs) above 1 TeV.  In addition, 
the VHE flux from 1ES\,1101$-$232, 
previously detected by HESS in 2004-2005, was monitored
during 2006 and 2007.}
   {As significant detections from the HESS observation program
are reported elsewhere, the results reported here 
are primarily integral flux upper limits.  
The average exposure for each of the 14 VHE-candidate
AGN is $\sim$7 h live time, and the observations have an
average energy threshold between 230 GeV and 590 GeV.
Upper limits for these 14 AGN range from $<$0.9\% to $<$4.9\% of the 
Crab Nebula flux, and eight of these are the most 
constraining ever reported for the object. 
The brief ($<$2.2 h each) large-zenith-angle observations yield 
upper limits for Mkn\,501 ($<$20\% Crab above 2.5 TeV) 
and 1ES\,1218+304 ($<$17\% Crab above 1.0 TeV), 
and a marginal detection ($3.5\sigma$) 
of Mkn\,421 (50\% Crab above 2.1 TeV). 1ES\,1101$-$232 was marginally 
detected ($3.6\sigma$, 1.7\% Crab above 260 GeV) during the 2006 (13.7 h
live time) observations,
but not in the 2007 (4.6 h live time) data. The upper limit in 2007 
($<$1.9\% Crab above 260 GeV) is below the average flux measured 
by HESS from 2004-2006.}
   {}


   \maketitle

\section{Introduction}

Active galactic nuclei (AGN) represent the only class of 
extragalactic objects known to emit VHE $\gamma$-rays.
These objects are found in the core of at least 5\%
of all galaxies.  Generally, they are characterized by very bright,
highly variable, non-thermal emission spanning the 
entire electromagnetic spectrum from radio waves to 
TeV $\gamma$-rays.  They are believed
to be powered by accretion of matter onto a super-massive 
($10^6$-$10^9$ solar mass) black hole.  
In the unified description of AGN (as reviewed in \cite{AGN_model}), 
this black hole is 
surrounded in the inner regions by an accretion disk, 
and in the outer regions by a thick torus of gas and dust.
Viewing these objects at various orientation 
angles with respect to the torus plane 
is believed to be the underlying cause of the
wide variations in their observed properties, and hence
numerous AGN classifications. 

In about 10\% of all AGN, collimated relativistic outflows of 
particles (known as relativistic jets)
exist, presumably along the magnetic field in the vicinity
of the black hole, and
approximately perpendicular to the accretion disk and torus plane.  
Of particular interest to high-energy astronomy is a class of 
radio-loud AGN known as blazars
which includes both  BL\,Lacertae (BL\,Lac) type objects
and Flat Spectrum Radio Quasars (FSRQ). For blazars, one 
relativistic jet is pointed close 
the observer's line-of-sight (i.e. towards Earth),
causing the observed emission to be relativistically
beamed.
Blazars are typically characterized by a double-peaked
broad-band SED, and for different
objects the peak energy of the lower or higher-energy 
components can differ by several orders of magnitude.
BL Lac objects are commonly categorized \cite{BLLacs} into three groups
depending on the position of the first peak:
low (LBL), intermediate (IBL), 
and high-frequency-peaked (HBL).
Essentially all\footnote{Out of the $\sim$20 VHE AGN, the only 
published exceptions are the radio galaxy M\,87 (\cite{HEGRA_M87,HESS_M87}) 
and BL\,Lacertae \cite{MAGIC_BlLac} (an LBL).
See Section~\ref{VHE_exception} for more details.}
AGN detected at VHE energies are HBL. 

The HESS array (\cite{HESS_jim, HESS_crab}) 
of four imaging atmospheric-Cherenkov 
telescopes located in Namibia is used to search for
VHE $\gamma$-ray emission from various
classes of astrophysical objects.
Approximately 30\% of the HESS observation program
is dedicated to studies of AGN, primarily blazars.  
Since 2005, the $\sim$300 hours per year of AGN observations are
divided between monitoring
known VHE-bright AGN for bright flaring episodes,
and searching for new VHE sources. 
For the discovery part of the AGN program, a candidate from 
a large, diverse sample of relatively nearby AGN is typically
observed for $\sim$10 hours.  If any of these observations 
show an indication for a signal (e.g., an excess with significance
more than $\sim$3 standard deviations), a deeper 
exposure is promptly scheduled to increase the overall significance
of the detection and to allow for a spectral measurement.
The detections resulting from 
the HESS AGN observation program are reported 
elsewhere (see Table~\ref{HESS_detections} for references). 
Results of HESS AGN observations taken from
January 2005 through July 2007, where 
VHE emission is not significantly detected, are presented here.

   \begin{table}
      \begin{minipage}[t]{\columnwidth}
      \caption{Ten AGN detected by HESS in order of redshift ($z$).}
\label{HESS_detections}
        \centering
        \renewcommand{\footnoterule}{}
         \begin{tabular}{c c c}
            \hline\hline
            \noalign{\smallskip}
		AGN\footnote{The asterisk denotes the seven AGN which were discovered 
	to emit VHE $\gamma$-rays by HESS.
	The dagger marks the objects with more than one HESS publication.} & $z$ & Reference\\
            \noalign{\smallskip}
            \hline
            \noalign{\smallskip}
	M\,87 & 0.004 & \cite{HESS_M87}\\
	Mkn\,421 & 0.030 & \cite{HESS_421}\\
	PKS\,0548-322$^{*}$ & 0.069 & \cite{HESS_0548}\\
	PKS\,2005-489$^{*}$ & 0.071 & \cite{HESS_2005}\\
	PKS\,2155$-$304$^{\dagger}$ & 0.116 & \cite{HESS_2155}\\
	1ES\,0229+200$^{*}$ & 0.139 & \cite{HESS_0229}\\
	H\,2356$-$309$^{* \dagger}$ & 0.165 & \cite{HESS_2356}\\
	1ES\,1101$-$232$^{* \dagger}$ & 0.186 & \cite{Gerd_1101}\\
	1ES\,0347$-$121$^{*}$ & 0.188 & \cite{HESS_0347}\\
	PG\,1553+113$^{*}$ & ? & \cite{HESS_1553}\\
	\noalign{\smallskip}
            \hline
       \end{tabular}
     \end{minipage}
   \end{table}

\section{Methodology}

The HESS AGN data presented here were generally taken in 
$\sim$28 minute data segments (runs) using {\it Wobble} 
mode, where the pointing of the array is slightly offset,
typically\footnote{For the only exception, BWE 0210+1159, 
the average offset is $\sim$0.9$^{\circ}$.} 
by $\pm$0.5$^{\circ}$ , from the position of the AGN.
For the results reported in this article, 
all HESS data passing the quality-selection criteria\footnote{Approximately
70\% of the relevant HESS AGN data pass these criteria.}
are processed using the standard HESS calibration \cite{calib_paper}
and analysis tools \cite{std_analysis}.  
The {\it standard cuts} \cite{std_analysis} are chosen
for the event-selection criteria.  On-source data
are taken from a circular region of radius $\theta_{cut}=0.11^{\circ}$,
appropriate for point-like sources, centered on each AGN. 
The {\it Reflected-Region} method \cite{bgmodel_paper}
is used to simultaneously estimate the cosmic-ray background 
(off-source data). Equation 17 in \cite{lima} is used
to calculate the significance (in standard deviations,
$\sigma$) of the observed excess.
All upper limits are determined following the method of \cite{UL_tech}.
The flux limits in Section~\ref{UL_ref} are all
calculated assuming a power-law spectrum
with photon index $\Gamma$=3.0 as none of these targets are 
previously detected in the VHE regime.
The limits are not very sensitive to the choice of photon index.
Assuming a moderately different photon index 
(i.e. $\Gamma$ between 2.5 and 3.5) 
changes the values by less than $\sim$10\%.  Even the choice of
an extremely soft spectrum ($\Gamma$=6.0) increases the limits
by only $\sim$25\%.
The fluxes and limits in Sections \ref{VHE_low_alt} and \ref{1101_Sect}
assume spectra previously measured for those sources.
All flux quantities and energy thresholds are corrected for decreases over time
of the absolute optical efficiency of the system, using efficiencies 
determined from simulated and observed muons (\cite{HESS_crab}).
Therefore the energy thresholds of these AGN observations are higher
than that of the newly commissioned HESS system
reported in Benbow (2005) as the 
total optical throughput is between $\sim$25\% to $\sim$35\% less
than initially measured in 2003.  The integral 
flux quantities given later are compared
to the HESS Crab Nebula flux calculated above the energy
threshold determined for each AGN respectively.
The reported percentages of the HESS Crab Nebula flux are 
calculated from the fit to HESS 
spectrum (\cite{HESS_crab}) and assume that the
function does not deviate from a power-law 
below the lowest energy measured ($\sim$500 GeV).

\section{HESS Observations of 14 AGN}
Approximately 30 AGN were observed by 
HESS from January 2005 through July 2007.
Some of these objects were previously shown by HESS
to emit VHE $\gamma$-rays, and the discoveries of VHE emission
from others are reported elsewhere. Of the remaining AGN
with non-zero good-quality exposure,
14 show no indication of any VHE emission.  Table~\ref{candidates} 
shows these 14 candidate AGN observed by HESS and the dates of
the observations. The total live time of the observations passing the 
standard data-quality selection criteria, and the mean zenith angle 
of those observations, are listed in
Table~\ref{upper_limits}.  The mean good-quality exposure for 
the candidates is 6.7 hours live time at a
mean zenith angle of 31$^{\circ}$. In 5 hours of observations, 
the sensitivity of HESS \cite{std_analysis}
enables a 5$\sigma$ detection of a $\sim$2\% Crab Nebula flux 
source at 20$^{\circ}$ zenith angle.  The sensitivity
of HESS decreases at larger zenith angles.  For example,
in 5 hours of observations at 45$^{\circ}$ zenith 
angle, a $\sim$3.5\% Crab Nebula flux source 
is detected at 5$\sigma$.

   \begin{table*}
      \begin{minipage}[t]{2.0\columnwidth}
      \caption{The candidate AGN ordered by right ascension in groups of
blazars and non-blazars.}
         \label{candidates}
        \centering
        \renewcommand{\footnoterule}{}
         \begin{tabular}{c c c c c c}
            \hline\hline
            \noalign{\smallskip}
	     Object\footnote{The coordinates (J2000), redshift, 
and type (HBL, IBL, LBL, FSRQ, Sy = Seyfert (I \& II),
FR II = Fanaroff-Riley II, NLS I = Narrow-line Seyfert I)
shown are taken from the SIMBAD Astronomical Database (http://simbad.u-strasbg.fr/simbad/) and the 
NASA/IPAC Extragalactic Database (http://nedwww.ipac.caltech.edu/).}\footnote{The superscript E denotes the
five candidates detected by the EGRET instrument aboard
the CGRO satellite (\cite{EGRET_AGN,EGRET_catalog,EGRET_new}).}
 & $\alpha_{\rm J2000}$ & $\delta_{\rm J2000}$ & $z$ & Type &  MJD$-$50000\footnote{Only the dates of the good-quality
HESS observations of each AGN are shown.} \\
             & [h m s] & [d m s] & & & \\
            \noalign{\smallskip}
            \hline
            \noalign{\smallskip}
	{\it Blazar} \\
            \noalign{\smallskip}
             III\,Zw\,2         & 00 10 31.0 & +10 58 30   & 0.0893 & FSRQ
		& 3944, 3953, 4267, 4270, 4272, 4274$-$76, 4278\\
	     BWE\,0210+116$^{\rm E}$	& 02 13 05.0 & +12 13 06   & 0.250  & LBL
	 	& 3966$-$69, 3971, 3974, 3976$-$78\\
	     1ES\,0323+022      & 03 26 14.0 & +02 25 15   & 0.147  & HBL
 		& 3668$-$69, 3676$-$78, 3998$-$4000\\
	     PKS 0521$-$365$^{\rm E}$	& 05 22 58.0 & $-$36 27 31 & 0.0553 & LBL
		& 4079$-$4081\\
	     3C\,273$^{\rm E}$ 		& 12 27 06.8 & +02 03 09 & 0.158 & FSRQ
		& 3411, 3413, 3494-95, 3497, 3499, 3502, 3505,\\
		& & & & & 4146$-$49, 4151, 4153, 4228$-$31\\
	     3C\,279$^{\rm E}$		& 12 56 11.2 & $-$05 47 22 & 0.536 & FSRQ
		& 4118$-$4121\\
             RBS\,1888          & 22 43 42.0 & $-$12 31 06 & 0.226 & HBL
		& 3914, 3916$-$18\\
	     HS\,2250+1926$^{\rm E}$ 	& 22 53 07.4 & +19 42 35 & 0.284 & FSRQ &4292$-$4302, 4304$-$05\\
             PKS\,2316$-$423    & 23 19 05.9 & $-$42 06 49 & 0.055 & IBL
		& 3919$-$23\\
 	     1ES\,2343$-$151    & 23 45 37.8 & $-$14 49 10 & 0.226 & IBL
		& 3592$-$95, 3597\\
\\	{\it Non-blazar} \\
            \noalign{\smallskip}
             NGC\,1068		& 02 42 40.8 & $-$00 00 48 & 0.00379 & Sy II
		& 4022, 4024, 4032\\
             Pictor\,A		& 05 19 49.7 & $-$45 46 45 & 0.0342 & FR II
		& 4051$-$54, 4056$-$57, 4060$-$64\\
	     PKS\,0558$-$504	& 05 59 46.8 & $-$50 26 39 & 0.137 & NLS I
		& 4110$-$4113, 4115$-$16, 4121\\
             NGC\,7469		& 23 03 15.8 & +08 52 26 & 0.0164 & Sy I
		& 4020$-$21, 4023$-$24, 4032\\
          \noalign{\smallskip}
            \hline
       \end{tabular}
     \end{minipage}
   \end{table*}

\subsection{The 14 Candidates\label{VHE_exception}}
A large majority of VHE-emitting AGN
are HBL, therefore these objects
are the primary targets of HESS AGN observation
program.  However, prominent examples of different
types of AGN are also observed.  Many of
the HBL observed by HESS have been detected, 
therefore the 14 candidates
discussed in this section are largely not HBL.

\subsubsection{Blazars}
Only two of the 14 candidates are HBL.  
The HBL 1ES\,0323+022 is 
the only target recommended 
by multiple authors (\cite{Stecker_AGN,perlman_AGN,luigi_AGN}) 
as a potential source of VHE $\gamma$-rays.
The remaining blazar targets include two 
IBL, five AGN (2 LBL and 3 FSRQ) 
detected by EGRET (\cite{EGRET_AGN,EGRET_catalog,EGRET_new}), 
and a high-frequency-peaked FSRQ. 
Although the lower-energy SED peaks of the other blazars 
might imply that they are less likely 
to emit VHE $\gamma$-rays, an LBL (the archetype BL\,Lacertae)
was recently detected \cite{MAGIC_BlLac} 
by the MAGIC Collaboration.

\subsubsection{Non-blazars}

Only one non-blazar AGN, hosted by the radio-loud Fanaroff-Riley (FR) I
galaxy M\,87, has been detected  (\cite{HEGRA_M87,HESS_M87})
at VHE energies. Of the four non-blazar targets, 
two (Pictor\,A and PKS\,0558$-$504)
are radio-loud objects and two (NGC 1068 and NGC 7649) are radio-weak. 
Pictor\,A, an FR II galaxy \cite{PicA}, is somewhat similar 
to a blazar, however its 
jet is oriented at a large angle with respect
to the line-of-sight from Earth. Therefore the VHE flux 
from Pictor A is naively expected to be low, 
as any emission would not be Doppler-boosted along the line of sight.  
PKS\,0558$-$504, a narrow-line Seyfert I galaxy, 
shows some evidence for beamed emission from a relativistic 
jet \cite{NLS1_ref}.
Detection of VHE emission from PKS\,0558$-$504 would strongly
support the presence of such a jet.  The radio-weak objects 
observed by HESS do not have relativistic jets.  They 
include the brightest and closest 
Seyfert II object (NGC 1068), and a Seyfert I (NGC 7469)
undergoing massive star formation near its 
nucleus \cite{starburst}.
It should be noted that VHE observations of non-blazar AGN
have the potential to detect VHE emission originating from sites 
other than jets (e.g., hot-spots in radio lobes) and
from mechanisms other than inverse-Compton scattering (e.g., 
proton acceleration in the vicinity of accreting black holes).

\subsection{Results\label{UL_ref}}

   \begin{table*}
      \begin{minipage}[t]{2.0\columnwidth}
      \caption{Results from HESS observations of 14 AGN.}
         \label{upper_limits}
        \centering
        \renewcommand{\footnoterule}{}
         \begin{tabular}{c c c c c c c c c c c c}
            \hline\hline
            \noalign{\smallskip}
	     Object\footnote{The superscript U marks the eight VHE upper limits (99.9\% c.l.)
that are the most constraining ever published for the corresponding
objects.} &
T & Z$_{\mathrm{obs}}$ & On & Off & Norm & Excess & S & E$_{\mathrm{th}}$ & I($>$E$_{\mathrm{th}}$) & Crab & P($\chi^2$)\\
		& [hrs] & [$^{\circ}$] & & & & & [$\sigma$] & [GeV] & [10$^{-12}$ cm$^{-2}$ s$^{-1}$] & \% & \\
            \noalign{\smallskip}
            \hline
            \noalign{\smallskip}
	{\it Blazars} \\
            \noalign{\smallskip}
             III\,Zw\,2$^{\rm U}$ & 4.9 & 38 & 169 & 1801 & 0.0916 & 4 & 0.3 & 430 & $<$2.14 &  $<$2.7 & 0.19\\
	     BWE\,0210+116$^{\rm U}$ & 6.0 & 43 & 176 & 3752 & 0.0504 & $-$13 & $-$0.9 & 530 & $<$0.72 &  $<$1.2 & 0.74\\
	     1ES\,0323+022 & 7.2 & 27 & 321 & 3302 & 0.0932 & 13 & 0.7 & 300 & $<$2.52 &  $<$1.9 & 0.59\\
	     PKS 0521$-$365$^{\rm U}$ & 3.1 & 26 & 180 & 1818 & 0.0928 & 11 & 0.8 & 310 & $<$5.40 &  $<$4.2 & 0.45\\
	     3C\,273$^{\rm U}$ & 16.5 & 29 & 848 & 8678 & 0.0932 & 39 & 1.3 & 300 & $<$1.97 &  $<$1.4 & 0.89\\
	     3C\,279$^{\rm U}$ & 2.0 & 26 & 100 & 1012 & 0.0942 & 5 & 0.5 & 300 & $<$3.98 &  $<$2.9 & 0.44\\
             RBS\,1888 & 2.4 & 15 & 184 & 1625 & 0.0949 & 30 & 2.2 & 240 & $<$9.26 &  $<$4.9 & 0.39\\
	     HS\,2250+1926$^{\rm U}$ & 17.5 & 44 & 597 & 6536 & 0.0923 & $-$6 & $-$0.2 & 590 & $<$0.45 &  $<$0.9 & 0.58\\
             PKS\,2316$-$423 & 4.1 & 20 & 299 & 2910 & 0.0929 & 29 & 1.6 & 270 & $<$4.74 &  $<$3.0 & 0.58\\
 	     1ES\,2343$-$151$^{\rm U}$ & 8.6 & 17 & 557 & 6286 & 0.0911 & $-$16 & $-$0.6 & 230 & $<$2.45 &  $<$1.2 & 0.67\\
	\\
	{\it Non-blazar} \\
            \noalign{\smallskip}
             NGC\,1068 & 1.8 & 29 & 75 & 687 & 0.0955 & 9 & 1.1 & 330 & $<$5.76 &  $<$4.9 & 0.47\\
             Pictor\,A & 7.9 & 31 & 397 & 4501 & 0.0932 & $-$23 & $-$1.1 & 320 & $<$2.45 &  $<$2.0 & 0.54\\
	     PKS\,0558$-$504$^{\rm U}$ & 8.3 & 28 & 426 & 4740 & 0.0929 & $-$14 & $-$0.7 & 310 & $<$2.38 &  $<$1.8 & 0.80\\
             NGC\,7469 & 3.4 & 34 & 98 & 1234 & 0.0909 & $-$14 & $-$1.3 & 330 & $<$1.38 &  $<$1.2 & 0.59\\
            \noalign{\smallskip}
            \hline
       \end{tabular}
     \end{minipage}
   \end{table*}

\begin{figure}
   \centering
      \includegraphics[width=8.7cm]{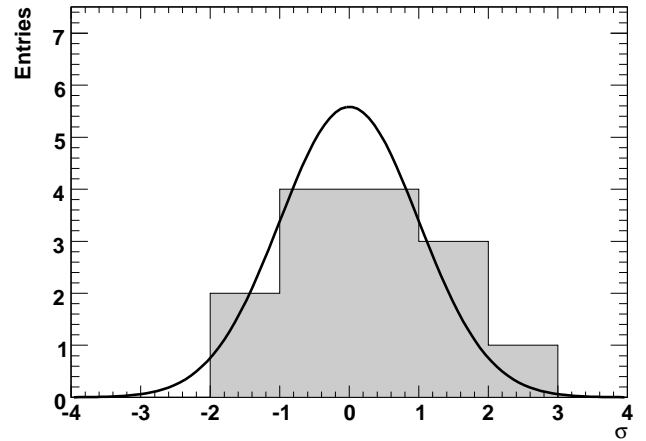} \\ [-0.3cm]
      \caption{Distribution of the significances observed 
from the 14 candidate AGN.  The curve
represents a Gaussian distribution with zero mean 
and a standard deviation of one.}
         \label{AGN_sigma}
   \end{figure}

No significant excess of VHE $\gamma$-rays is found 
from any of the 14 AGN in the given exposure time.
The total good-quality live time (T), 
mean zenith angle of observation (Z$_{\mathrm{obs}}$),
number of on-source and off-source counts, 
off-source normalization, observed
excess, and significance (S) of the excess in standard deviations
are given for each of the 14 AGN in 
Table~\ref{upper_limits}.  Figure~\ref{AGN_sigma} shows the distribution 
of the significance observed from the direction of each of the 14 AGN.  
The distribution is slightly skewed
towards positive values. However, combining the excess from 
all 14 candidates only yields a total of 54 events 
and a statistical significance of
0.8$\sigma$.  A search for serendipitous 
source discoveries in the HESS field-of-view centered on each of the 
AGN also yields no significant excess.

Table~\ref{upper_limits} shows 99.9\% confidence level (c.l.)
upper limits on the integral flux (I)
above the energy threshold of the observations (E$_{\mathrm{th}}$) 
and the corresponding percentage of the HESS Crab Nebula flux.  
The systematic error on a HESS integral flux measurement
is estimated to be $\sim$20\%, and it is not included in the
calculation of the upper limits.

A search for VHE flares from each observed AGN was also performed. Here
the nightly integral flux above the average energy threshold 
was calculated from the observed excess
and fit by a constant.  Table~\ref{candidates} shows the dates each 
AGN was observed and Table~\ref{upper_limits} shows
the resulting $\chi^2$ probability P($\chi^2$).
As each  $\chi^2$ probability is acceptable, no evidence for 
any VHE flares is found.  

\section{Low Altitude Observations of VHE AGN\label{VHE_low_alt}}

As HESS is located in the Southern Hemisphere, 
AGN with declinations greater than $30^{\circ}$
culminate at low altitudes ($<37^{\circ}$)
where the energy threshold of HESS is considerably higher 
and the sensitivity is reduced.  However, due to the 
large effective area of the HESS array at low altitudes,
observations of northern AGN probe higher energies than
typically studied by Northern Hemisphere instruments.
Simultaneous observations of the same target by both
Northern and Southern Hemisphere instruments sample
different parts of the VHE spectrum, potentially
increasing the spectral coverage to several orders of
magnitude and allowing for cross-calibration between the 
detectors (see, e.g., \cite{HESS_MAGIC_421}).  
Therefore, three northern AGN, known to emit VHE $\gamma$-rays,
were briefly (good-quality live time $<$2.2 h) 
observed at low altitudes with HESS.  For
two of these targets (1ES\,1218+304 and Mkn\,501)
simultaneous observations
were performed by the MAGIC VHE telescope
and the Suzaku X-ray satellite \cite{Suzaku_info}.

\subsection{Mkn\,421}
HESS observed the well-known VHE emitter \cite{mkn421_disc}
Mkn\,421 on April 12, 2005 from
21:01 to 22:00 UTC.  Both of the two 28-minute
runs pass the standard quality selection
criteria, yielding a data set of 0.9 h live time at a 
mean zenith angle of $63^{\circ}$.
A total of 81 on-source events, 873 off-source events,
with an on-off normalization $\alpha$=0.0607 are measured.
A marginal excess (28 events, 3.5$\sigma$)
is found.  The corresponding integral flux
above the 2.1 TeV analysis threshold is  
I($>$2.1 TeV) = $(3.4\pm1.2_{\rm stat}\pm0.7_{\rm syst}) \times 10^{-12}$
cm$^{-2}$\,s$^{-1}$, or 50\% of the HESS Crab Nebula flux. 
The flux in each run differs by only 3\%, 
clearly consistent within statistical errors.
As the spectrum of Mkn\,421 above 2 TeV is known to 
vary along with its flux (\cite{HESS_421}),
the flux calculated here assumes
the time-averaged spectrum,
a power law with photon index $\Gamma=2.39$ and 
an exponential cut-off
at 3.6 TeV, measured (\cite{HEGRA_421}) up to 10 TeV 
by HEGRA between December 1999 and May 2000 during a state
of comparable flux. 

\subsection{1ES\,1218+304}
1ES\,1218+304, discovered as a VHE emitter by
the MAGIC collaboration \cite{MAGIC_1218}, 
was the target of a HESS observation
campaign on May 19$-$21, 2006.  
All data from the first two nights of observations fail
the data-quality criteria due to bad weather.  On
the last night, a total of five good-quality observation
runs were taken between 17:43 and 21:05 UTC
yielding a data set of 1.8 h live time at a 
mean zenith angle of $56^{\circ}$.
One of the telescopes was not functioning properly during
the data taking and is excluded from the analysis, 
resulting in a slightly reduced sensitivity.
A total of 61 on-source events and 
590 off-source events ($\alpha$=0.0880)
pass the event selection criteria.
The resulting excess is not significant (9 events, 
1.2$\sigma$). Assuming $\Gamma$=3.0, as measured
\cite{MAGIC_1218} by MAGIC from $\sim$100 GeV 
to $\sim$600 GeV, the 99.9\% c.l. limit on the
integral flux above the 1.0 TeV analysis threshold is  
I($>$1.0 TeV) $ < (3.9\pm0.8_{\rm syst}) \times 10^{-12}$
cm$^{-2}$\,s$^{-1}$.  This corresponds to
17\% of the HESS Crab Nebula flux.  The upper limit is
$\sim$6 times higher than the flux above 1 TeV determined
from an extrapolation of the MAGIC spectrum.

\subsection{Mkn\,501}
HESS observations of Mkn\,501 \cite{mkn501_disc}
occurred on July 18, 2006 from
18:49 to 21:18 UTC.  All five 28-minute runs
pass the standard quality-selection
criteria, yielding a data set of 2.2 h live time at a 
mean zenith angle of $64^{\circ}$.
A total of 112 on-source events and 1328 off-source events 
($\alpha$=0.0915) are measured.
Mkn\,501 is not detected by HESS
as the resulting excess is $-9$ events ($-0.8$$\sigma$).
Assuming $\Gamma=2.6$ as measured \cite{HEGRA_501}
above 1.5 TeV by HEGRA,
the limit (99.9\% c.l.) on the
integral flux above the 2.5 TeV analysis threshold is  
I($>$2.5 TeV) $ < (1.0\pm0.2_{\rm syst}) \times 10^{-12}$
cm$^{-2}$\,s$^{-1}$, or 20\% of the 
HESS Crab Nebula flux.

\section{VHE Monitoring of 1ES\,1101$-$232\label{1101_Sect}}
1ES\,1101$-$232 was discovered by HESS (\cite{Nature_EBL,Gerd_1101}) 
to emit VHE $\gamma$-rays during observations in 2004-2005.
As part of a campaign to monitor its VHE flux, it was re-observed 
for 17.4 h between December 29, 2005 and May 28, 2006, and for 6.4 h 
from April 9$-$17, 2007.  A total of 18.3 h,
13.7 h in 2006 and 4.6 h in 2007, pass the
data-quality criteria. In addition, the Suzaku X-ray satellite   
observed 1ES\,1101-232 from 16:07 UTC on May 25, 2006 
until 05:11 UTC on May 27, 2006,
with an average observation efficiency of $\sim$47\%.
(Suzaku Observation Log: http://www.astro.isas.ac.jp/suzaku/index.html.en).
A total of 4.3 h of good-quality HESS data are simultaneous 
to the Suzaku observations.
The results from the HESS observations 
are given in Table~\ref{1101_results} for the various observation epochs.
The number of on-source and off-source counts, 
the off-source normalization, the observed
excess, the significance (S) of the excess in standard deviations,
and the integral flux above the analysis threshold of 260 GeV are shown
in the table.  The object is marginally detected in the 2006 observations, as well
as during the Suzaku epoch.  1ES\,1101$-$232 is not detected in
2007. All flux quantities in Table~\ref{1101_results} assume
the photon index $\Gamma=2.94$  previously measured 
(\cite{Gerd_1101}) by HESS for 1ES\,1101$-$232.

   \begin{table*}
      \begin{minipage}[t]{2.0\columnwidth}
      \caption{Results from HESS observations of 1ES\,1101$-$232
during various observation epochs.}
         \label{1101_results}
        \centering
        \renewcommand{\footnoterule}{}
         \begin{tabular}{c c c c c c c}
            \hline\hline
            \noalign{\smallskip}
		Epoch & On & Off & Norm & Excess & S & I($>$ 260 GeV)\footnote{The systematic error
on the integral flux above 260 GeV is 20\%. The 99.9\% confidence limit
is given for the 2007 observations.}
\\
		& & & & & [$\sigma$] & [$10^{-12}$ cm$^{-2}$ s$^{-1}$]\\
            \noalign{\smallskip}
            \hline
            \noalign{\smallskip}
		2006 & 1016 & 9803 & 0.0917 & 117 & 3.6 & 2.8$\pm$0.7$_{\rm stat}$\\
		2007 & 309  & 3220 & 0.0909 & 16  & 0.9 & $<$3.1\\
            \noalign{\smallskip}
            \hline
            \noalign{\smallskip}
		Total & 1325 & 13023 & 0.0915 & 133 & 3.6 & 2.0$\pm$0.6$_{\rm stat}$\\
	\noalign{\smallskip}
            \hline
            \noalign{\smallskip}
            Suzaku & 323 & 2938 & 0.0927 & 51 & 2.9 & 3.2$\pm$1.4$_{\rm stat}$\\
	\noalign{\smallskip}
            \hline
       \end{tabular}
     \end{minipage}
   \end{table*}

   \begin{figure}
   \centering
      \includegraphics[width=8.7cm]{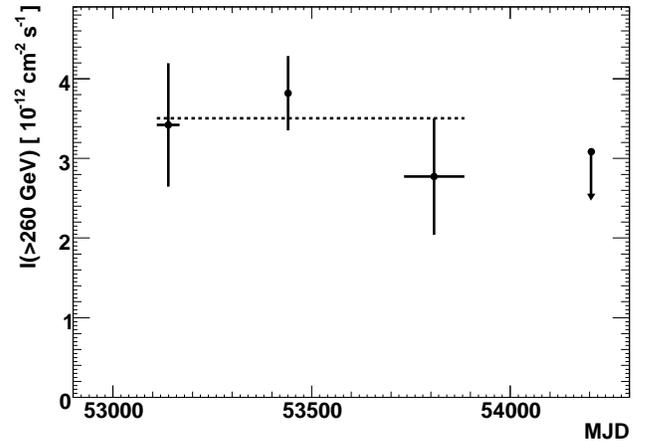} \\ [-0.3cm]
      \caption{Integral flux ($>$260 GeV) measured by HESS
from 1ES\,1101$-$232 in annual bins.  The 2004 and 2005 data
were previously published (\cite{Gerd_1101}).  The horizontal
error bars reflect the actual observation dates (first to last
observation in each period).  The dashed line
is the average flux measured from 2004-2006.  The upper limit
in 2007 is at the 99.9\% confidence level.}
         \label{1101_lc}
   \end{figure}

There is no indication of significant variability in the 
VHE flux from 1ES\,1101$-$232 within the 2006 or 2007 data,
as the nightly flux within each year is well fit by a constant.
Figure~\ref{1101_lc} shows a light curve
of the integral flux above 260 GeV in annual bins 
for the observations here, as well as those previously 
published for the 2004-2005 data (\cite{Gerd_1101}).  
The 99.9\% confidence limit on the integral flux in 2007 is 
marginally inconsistent with the average value, 
I($>$260 GeV) = $(3.50\pm0.35) \times 10^{-12}$ cm$^{-2}$ s$^{-1}$,
measured by HESS from 2004-2006.  

\section{Discussion}

One of the defining characteristics of AGN is their extreme
variability. The VHE flux from any of these AGN may increase 
significantly during future flaring episodes 
(see, e.g., \cite{2155_flare}) and could potentially
exceed the limits presented here.  In
addition, accurate modeling of the SED 
requires that the state of the source is accounted for.  
Therefore it must be emphasized that the upper limits 
reported here constrain the flux of the AGN only 
during the observation time (see Table~\ref{candidates}
for the 14 candidates). The simultaneous Suzaku X-ray data 
from Mkn\,501, 1ES\,1218+304, and 1ES\,1101$-$232, make the
HESS results from these objects particularly useful.
In the absence of simultaneous observations at lower
energies, it is recommended that the limits 
be conservatively interpreted as limits 
on the steady-component or quiescent flux from the AGN. 

Interpretation of the SED of an AGN not only requires accounting
for the state of the source, but also the redshift and energy dependent
absorption (\cite{EBL_1}) 
of VHE photons on the Extragalactic Background Light (EBL). 
The EBL density (\cite{EBL_2,Nature_EBL}), 
and hence optical depth ($\tau$),
is only roughly constrained at wavelengths 
relevant to the HESS measurements.  Therefore the effect
of the EBL on the HESS limits is quantified with a maximal
and minimal EBL density (i.e. maximal and minimal EBL absorption).
Here the EBL parameterization of \cite{Primack05_EBL} is used.  
This EBL is near the lower limits on the EBL density 
provided by galaxy counts \cite{Madau} and is considered
a minimal EBL scenario.  Scaling the normalization
of the minimal EBL parameterization
by a factor of 1.6 yields a density approximately
at the level of the most-constraining upper limits (\cite{Nature_EBL})
derived from VHE observations of 1ES\,1101$-$232, and is therefore
used here as a maximal EBL parameterization.
Table~\ref{EBL_table} shows the HESS integral flux 
limits from Table~\ref{upper_limits} converted
to a limit on the observed differential flux ($F_{\rm obs}$)
at the energy threshold of the HESS measurement,
as well as scaling factors, $A=e^{\tau}$, to convert each to an intrinsic 
flux limit (i.e. correct for absorption) 
for both the minimal and maximal EBL scenarios.
The corresponding information for the HESS limits on 1ES\,1218+304
and Mkn\,501 are also shown in Table~\ref{EBL_table}.
For Mkn\,421 (z=0.030) the differential flux
at 2.1 TeV is $F_{\rm obs}$ = 
$(5.7\pm1.9_{\rm stat}\pm1.1_{\rm syst}) \times 10^{-12}$ 
cm$^{-2}$ s$^{-1}$ TeV$^{-1}$, and the minimal and maximal
scaling factors are  $A_{\rm Min} = 1.36$ and  $A_{\rm Max} = 1.63$,
respectively. During the Suzaku epoch for 1ES\,1101$-$232 (z=0.186) 
the differential flux at 0.26 TeV is $F_{\rm obs}$ = 
$(2.4\pm1.0_{\rm stat}\pm0.5_{\rm syst}) \times 10^{-11}$ 
cm$^{-2}$ s$^{-1}$ TeV$^{-1}$, and the minimal and maximal
scaling factors are  $A_{\rm Min} = 1.40$ and  $A_{\rm Max} = 1.72$,
respectively.  The differential flux quantities 
for the three other 1ES\,1101$-$232 epochs
can be calculated using the ratios of the integral fluxes
reported in Table~\ref{1101_results}.

   \begin{table*}
    \begin{minipage}[t]{2.0\columnwidth}
      \caption{Differential flux upper limits from the HESS observations.}
         \label{EBL_table}
        \centering
        \renewcommand{\footnoterule}{}
         \begin{tabular}{c c c c c c}
            \hline\hline
            \noalign{\smallskip}
	     Object & $z$ & E$_{\mathrm{th}}$ & $F_{\rm obs}$\footnote{The observed ($F_{\rm obs}$) 
differential flux upper limits at the energy threshold of the HESS observations 
(E$_{\mathrm{th}}$) are calculated at the 99.9\% confidence level. 
The limit on the intrinsic differential flux is given by multiplying 
the observed limit by the scaling factor, $A=e^{\tau}$, for either a minimal ($A_{\rm Min}$) or
maximal ($A_{\rm Max}$) EBL scenario.} 

& $A_{\rm Min}$ & $A_{\rm Max}$\\
		& & [TeV] & [ 10$^{-11}$ cm$^{-2}$ s$^{-1}$ TeV$^{-1}$] & &\\
	    \noalign{\smallskip}	
            \hline
            \noalign{\smallskip}
	{\it VHE blazars} \\
            \noalign{\smallskip}
	1ES\,1218+304 & 0.182 & 1.0 & $<$0.78 & 4.29 & 10.3\\
	Mkn\,501 & 0.0336 & 2.5 &  $<$0.064 & 1.43 & 1.76\\
\\
	{\it Blazars} \\
            \noalign{\smallskip}
             III\,Zw\,2 & 0.0893 & 0.43 &  $<$1.00 & 1.34 & 1.61\\
	     BWE\,0210+116 & 0.250 & 0.53 &  $<$0.27 & 3.39 & 7.06\\
	     1ES\,0323+022    & 0.147 & 0.30 &  $<$1.68 & 1.37 & 1.65\\
	     PKS 0521$-$365 & 0.0553 & 0.31 &  $<$3.48 & 1.12 & 1.20\\
	     3C\,273 & 0.158 & 0.30 &  $<$1.31 & 1.40 & 1.72\\
	     3C\,279 & 0.536 & 0.30 &  $<$2.65 &  5.12 & 13.7\\
             RBS\,1888        & 0.226 & 0.24 &  $<$7.72 & 1.46 & 1.83\\
	     HS\,2250+1926 & 0.284 & 0.59 &  $<$0.15 & 4.88 & 12.6\\
             PKS\,2316$-$423    & 0.055 & 0.27 &  $<$3.51 & 1.09 & 1.16\\
 	     1ES\,2343$-$151    & 0.226 & 0.23 &  $<$2.13 & 1.42 & 1.76\\
	\\
	{\it Non-blazar} \\
            \noalign{\smallskip}
             NGC\,1068        & 0.00379 & 0.33 &  $<$3.49 & 1.01 & 1.01\\
             Pictor\,A        & 0.0342 & 0.32 &  $<$1.53 & 1.07 & 1.12\\
	     PKS\,0558$-$504    & 0.137 & 0.31 &  $<$1.54 & 1.35 & 1.62\\
             NGC\,7469        & 0.0164 & 0.33 &  $<$0.84 & 1.04 & 1.06\\
            \noalign{\smallskip}
            \hline
       \end{tabular}
     \end{minipage}
   \end{table*}

\section{Conclusions}

HESS observed a large sample of AGN between January 2005 and July
2007 as part of a campaign to identify new VHE-bright AGN.
Results presented here detail the observations
of 14 candidates for which no significant excess was found.
The corresponding upper limits on the VHE flux are the most 
stringent to date for eight of the candidates, and are
only surpassed by those from earlier HESS 
observations \cite{HESS_AGN_UL} for the other six candidates. 
In addition the results from HESS observations of four AGN
known to emit VHE $\gamma$-rays are presented.  Although
only a marginal excess is measured for two of these objects,
and no excess is observed from the other two, the presence of
simultaneous X-ray data for three of these objects enables
more accurate SED modeling than possible with archival data.

With the detection of ten VHE AGN, including the discovery of
seven, the HESS AGN observation program has been highly successful. 
However, despite more than five years of operation, the observation 
program is not complete as many proposed candidates 
have either not yet been observed
or only have a fraction of their intended exposure.  
Therefore, the prospects of finding additional VHE-bright AGN 
with HESS are still excellent.

\begin{acknowledgements}

The support of the Namibian authorities and of the University of Namibia
in facilitating the construction and operation of H.E.S.S. is gratefully
acknowledged, as is the support by the German Ministry for Education and
Research (BMBF), the Max Planck Society, the French Ministry for Research,
the CNRS-IN2P3 and the Astroparticle Interdisciplinary Programme of the
CNRS, the U.K. Science and Technology Facilities Council (STFC),
the IPNP of the Charles University, the Polish Ministry of Science and 
Higher Education, the South African Department of
Science and Technology and National Research Foundation, and by the
University of Namibia. We appreciate the excellent work of the technical
support staff in Berlin, Durham, Hamburg, Heidelberg, Palaiseau, Paris,
Saclay, and in Namibia in the construction and operation of the
equipment.  This research has made use of the SIMBAD database, 
operated at CDS, Strasbourg, France. This research has also made use of the 
NASA/IPAC Extragalactic Database (NED) which is operated by the Jet Propulsion Laboratory, 
California Institute of Technology, under contract with the National Aeronautics and Space Administration.

\end{acknowledgements}

\end{document}